\documentstyle[preprint,aps]{revtex}

\newcommand{\beq}{\begin{equation}}
\newcommand{\eeq}{\end{equation}}

\newcommand{\cavo}{$\rm{CaV}_4{\rm O}_9$}

\newcommand{\biup}{b_{i \uparrow}^{\dagger}}
\newcommand{\bidp}{b_{i \downarrow}^{\dagger}}
\newcommand{\bjup}{b_{j \uparrow}^{\dagger}}
\newcommand{\bjdp}{b_{j \downarrow}^{\dagger}}
\newcommand{\biu}{b_{i \uparrow}}
\newcommand{\bid}{b_{i \downarrow}}
\newcommand{\bju}{b_{j \uparrow}}
\newcommand{\bjd}{b_{j \downarrow}}

\newcommand{\Aij}{A_{i j}}
\newcommand{\Aijp}{A_{i j}^{\dagger}}
\newcommand{\Bij}{B_{i j}}
\newcommand{\Bijp}{B_{i j}^{\dagger}}

\begin{document}

\bibliographystyle{plain}

\title{Spin gap in \cavo: a large S approach}
\vskip0.5truecm
\author {{\sc Marc Albrecht} and {\sc Fr\'ed\'eric Mila}}
\vskip0.5truecm
\address{Laboratoire de Physique Quantique\\
Universit\'e Paul Sabatier\\
31062 TOULOUSE \\
FRANCE}
\maketitle

\begin{abstract}
We study the stability of the 4-site plaquette order in a model recently
introduced by Katoh and Imada to explain the spin gap observed in \cavo. We
show that all relevant types of order (4-site plaquette, dimerized, N\'eel)
can be described within the framework of Schwinger boson mean field theory,
and we make predictions regarding their relative stability. The results are
compared to exact diagonalization of finite clusters.
\end{abstract}

\newpage


In a recent paper, Taniguchi {\em et al.}~\cite{taniguchi}
reported the presence of a spin
gap in \cavo.
The structure of the two dimensional crystal of $\rm{VO}_4$
pyramids of \cavo\
is shown on figure~\ref{reseau}. In
the Mott insulating phase, one can consider Vanadium atoms
as localized spin one half coupled by superexchange interactions,
and, as pointed out by Katoh and Imada~\cite{katoh} a minimal model consists
of spin $1/2$ antiferromagnetic Heisenberg model including an exchange term
$J$ between nearest neighbors Vanadium atoms. These authors have shown
using Monte Carlo simulations that this model has a spin gap. In order
to determine the origin of the gap, they introduce a slightly more general
Hamiltonian:

\beq H=J \sum_{<\!ij\!>} \vec{S_i}.\vec{S_j}
      +J'\sum_{<\!ij\!>'} \vec{S_i}.\vec{S_j} \eeq
where $<\!ij\!>$ (resp. $<\!ij\!>'$) means full line (resp. dashed line)
links on figure~\ref{reseau}. Using first order perturbation theory in
$J'/J$, they argue that the system keep the same type of order as for $J'=0$
{\em i.e.} a product of singlet on each plaquette.

In fact, introducing two links
gives us a new
quantum spin model which might have interesting phase transitions:
For $J \gg J'$ the system is gapped and the ground state is the
product of the singlet state on each square plaquette while for $J'\gg J$
the origin of the spin gap is quite different. The ground state is then
the product of singlets on each inter-plaquette bonds. The question that
naturally arises
concerns the way the system is passing from one type
of order to the other while increasing $J'/J$.
Another question that can also be asked is whether
N\'eel long range order can exist for $S=1/2$ -- the system
is not frustrated! -- and,
if this is the case, what is  the nature of the
transition between this ordered state and the previously mentioned
dimerized states.

In this Rapid Communication, our aim  is to answer these questions
using a large S approach.
A linear spin wave (LSWT) calculation was done very recently by Ueda
{\em at al.}~\cite{ueda}. This theory predicts that the system is
N\'eel ordered for a large region of parameters: $.25 < J'/J < 6 $.
As these authors say, such the bounds seem to be incompatible with
second order perturbation theory in $J'/J$ and $J/J'$.
This could  be explained by the tendency LSWT
to favor N\'eel order. Still, the perturbation theory is not expected to be
reliable around $J'=J$, and the possibility of stabilizing N\'eel order in
that region remains.


Another  method to check the presence of N\'eel order in the context of a
large S approach
is to use the Schwinger boson
mean field theory (SBMFT)~\cite{arovas}.
One advantage is that it is a improvement over LSWT because it includes
higher order corrections in a self-consistent way. But more importantly,
this method is able to described
not only  the N\'eel ordered phase but also  the two types of
valence bond order which are present in the limits $J\rightarrow 0$ and
$J'\rightarrow 0$.
This method starts from a representation of the spin algebra in terms of
bosonic operators: $ \vec{S_{i}} = \frac{1}{2}
b_{i\sigma}^{\dagger}  \vec{\sigma}_{\sigma \sigma'} b_{i\sigma'} $, the size
of the spin being fixed by a constraint on the number of particles on each
site:
$b_{i\uparrow}^{\dagger} b_{i\uparrow}  + b_{i\downarrow}^{\dagger}
b_{i\downarrow} =2S$.
Defining operators that are quadratic in terms of the bosonic operators by
 $2\Bijp = \biup \bju + \bidp \bjd $ and $2\Aijp = \biup \bjdp
- \bidp \bjup $, the Hamiltonian can be written:
\beq
H = \sum_{(i,j)} J_{ij}(:\Bijp \Bij: - \Aijp \Aij)
\eeq
At the mean-field level, one introduces the following order parameters:
$ <\!\Aijp\!> = 2 \alpha_{ij} $ and $ <\!\Bijp\!>  =  2 \beta_{ij} $
and the Hamiltonian is replaced with:
\beq
H_{MF}= \sum_{(i,j)} J_{ij}\left( \beta_{ij}  (\Bij+\Bijp)
                          - \alpha_{ij} (\Aij+\Aijp)
                            -\beta_{ij}^2 +\alpha_{ij}^2 \right)
\eeq
Finally, the local constraint is replaced by a global one and is
enforced only on the average through the addition to the Hamiltonian of a term
$ \mu \sum_{i} \left( \biup \biu + \bidp \bid -2S \right) $,
where the chemical potential $\mu$ plays the role of a
Lagrange parameter.
Because there are only antiferromagnetic like correlations in this model,
we have: $ <\Bij>=0 $ for all pairs of spin $(i,j)$.

These equations have two types of solution: In the first one, the dispersion
relation is vanishing at three points in the Brillouin zone.
This means that the bosons are condensed with
a macroscopic expectation $S^{*}$ for the occupation number of the
zero energy modes with the consequence that
\beq \lim_{|i-j| \rightarrow \infty}
    <\vec{S_{i}}.\vec{S_{j}}> = \pm {S^{*}}^2  \eeq
The second kind of solution has a gap in the excitation spectra.
There is then no long range order. The mean field equations have
two remarkable solutions of this type: one has the same energy and the
same correlation functions as the plaquette-singlet one, while the
other as the same physical properties as the dimerized one.

The phase diagram of the model under investigation is represented on
figure~\ref{phase}. At the SBMFT approximation, there is only one
solution for a given value of $J'/J$. For $J'/J < 0.6 $ we find a
gapped solution with an energy per site $-.5J$: this solution has
the same correlation functions as the plaquette-singlet.
At the critical value  $J'/J \approx .6$ the gap closes.
For $.6 < J'/J < 2.4 $ the system exhibits N\'eel
long range order. At $J'/J = 2.4$ a gap opens again and the order
corresponds to dimers on the $J'$ bonds.
The order parameters $ <\!\alpha_{ij}\!>$
are continuous through both transitions which shows they are second order.
Note that this result is meaningful and not a mere consequence of the
approximation used. For another model, using SBMFT, we found a first order
transition between LRO and valence bond order~\cite{albrecht}.


We have also done exact diagonalisation of small clusters in order
to check our results and to have an estimation of the gap
in the spin excitations. As there are four atoms per unit cell,
one can only study square clusters having $4N$ atoms.
Besides if $N$ is odd, periodic
boundary conditions are frustrating.
So the
only tractable clusters have 8 and 16 spins. The next
cluster (24 sites) dose not have the symmetry of the square lattice.
For these two
clusters, we have calculated
the ground state energy, the spin gap $\Delta_N$ and the staggered
magnetization defined by $ M^* = < \phi | (1/N \sum_i \epsilon_i
S_{i}^z )^2 | \phi > $ where $|\phi>$ is the ground state of the system
and $\epsilon_i= \pm 1$ depending on the orientation of the corresponding
classical spin in the N\'eel state.

The different quantities can be scaled to their thermodynamic
limit~\cite{neuberger}:

\beq \frac{E_N}{N} = E_0 +  \frac{C_1}{N^{3/2}} \eeq

\beq \Delta_N = \Delta_0 + \frac{C_2}{N} \eeq

\beq M_{N}^{*} =M^{*}_{0} + \frac{C_3}{N^{1/2}} \eeq

The results are displayed on figures~\ref{energy},\ref{phase} and
\ref{gap}.
Taken literally, the results are inconsistent because the magnetization and
the gap cannot be non zero at the same time. But given the sizes we could
look at, very small values of $M^*$ or $\Delta$ are not accessible and only
sizable values should considered as meaningful. With this proviso,
we find a qualitative agreement between exact diagonalisation and SBMFT.
The staggered magnetization we deduced from
finite size extrapolation is rather large for $.6 < J'/J < 2.4 $ and
comparable to the value calculated within SBMFT.
In that region of parameters, the value of the extrapolated gap
is small and reaches a minimum of $0.04J$ at $J'/J=1.25$.
This two facts suggests that the SBMFT calculation are reliable, and
that the system under investigation sustains N\'eel long range order for $ J'/J
\approx 1.25$

To summarize, the plaquette
singlet order proposed by Katoh and Imada to explain the magnetic properties
of \cavo\ can be found in the context of $1/S$ approach provided one goes
beyond LSW theory by using SBMFT \cite{note}).
This method enables one to study the stability of this type of order with
respect to the switching on of a coupling $J'$ between the plaquettes. The
result is that N\'eel order is recovered in  a sizable region around $J'/J
\approx 1.25$, to disappear again when $J'/J$ is large enough in favor of a
dimerized order. It would be very interesting to see if Monte Carlo
calculations of the gap sustain the presence of a gapless region.

At a more quantitative level, let us note that the predictions of SBMFT
disagree with the results of Katoh and Imada, who reported a gap of about
$0.1J$ for $J'=J$, while the SBMFT predicts the system to be already gapless
for $J'/J\approx 0.6$. A possible explanation is that there is indeed an
ordered phase but it is smaller then what SBMFT predicts, and in particular
it does not include the point $J'=J$.
Note however that, contrary to Katoh and Imada's results,
very recent Monte Carlo simulations by Troyer {\em et al.}~\cite{troyer}
suggest that N\'eel order is present for the case $J'=J$.
More work is needed to check that
point. In any case, as far as \cavo\ is concerned, this might be of minor
importance. As pointed out recently by Sano {\em et al.}~\cite{sano},
the plaquette order can be stabilized by introducing some frustration into
the system.

We thank the authors of reference~\cite{ueda}, and especially H. Kontani,
for pointing out to us a mistake in the first version of this manuscript. We
also thanks Matthias Troyer for informing us of his results prior to
publication.
The numerical calculations reported here were possible thanks to computing
time made available by IDRIS, Orsay (France).




\begin{figure}
\caption{ (a) Schematic structure of the ${\rm VO}_4$
pyramids plane. (b) effective Heisenberg model }
\label{reseau}
\end{figure}

\begin{figure}
\caption{Ground state energy $J'/J$ for $N=8$ $(\circ)$ and $N=16$ $(\diamond)$
clusters and values extrapolated for $N=\infty$ $(\bullet)$ together with
SBMFT energy (full line)}
\label{energy}
\end{figure}

\begin{figure}
\caption{Magnetization versus $J'/J$ for $N=8$ $(\circ)$ and $N=16$
$(\diamond)$
clusters and values extrapolated for $N=\infty$ $(\bullet)$ together with
SBMFT energy (full line)}
\label{phase}
\end{figure}

\begin{figure}
\caption{Spin gap versus $J'/J$ for $N=8$ $(\circ)$ and $N=16$ $(\diamond)$
clusters and values extrapolated for $N=\infty$ $(\bullet)$ }
\label{gap}
\end{figure}

\end{document}